# Radiation hardened beam instrumentations for multi-Mega-Watt beam facilities

Edited by Katsuya Yonehara (Fermi National Accelerator Lab.)

The radiation hardened beam instrumentations workshop was held on February 3rd and 4th, 2022 to identify the technological challenges on the beam instrumentation for multi-Mega-Watt (MW) beam facilities [1]. We invited a speaker from a high energy neutrino beam facility (CERN, Fermilab and J-PARC), a collider detector group (CERN and Fermilab), a neutron spallation source facility (ESS, J-PARC and SNS), and a rare-isotope beam facility (FRIB). Although, the operational beam parameter for each institution is unique due to their different physics goals (see Table 1), we realize that there is a common issue on the beam instrumentation. Therefore, it will be tackled by a collaborative R&D effort among facilities. We summarize an output from the workshop in a following paragraph and Table 2, and propose a possible R&D for the future radiation hardened beam instrumentations with a global effort.

Table 1: Beam parameter reported in the workshop. A value in a parenthesis is a future plan.

| *Institution* | *Beam energy* | *Beam intensity* | *Beam power* | *Radiation dose at beam instrumentations* |
|---|---|---|---|---|
| CERN CNGS | 400 GeV | 2.4e13 protons per pulse | 500 kW | |
| CERN LHC | 7 TeV | 1.2e11 protons per bunch, 2808 bunches per beam | 362 MJ per beam | |
| Fermilab NuMI | 120 GeV | 5.5e13 (6.5e13) per 1.2 sec | 870 kW (1 MW) | $10^{10}$ mSv/hr at target $10^4$ mSv/hr at muon monitor |
| Fermilab LBNF | 60-120 GeV | | 1.2-2.4 MW | |
| J-PARC neutrino | 30 GeV | 3.2e14 protons per 2.48 sec (1.32 sec) | Present 515 kW (700 kW) | |
| ESS | 2 GeV | | | |
| J-PARC MLF | 3 GeV | (4.5e8 pulses, 25 Hz) | 730 kW (1 MW) | 30 MGy/6 y |
| SNS | 1 GeV (1.3 GeV) | 1.5e14 per beam | 1.4 MW (2 MW) | |
| FRIB | 200 MeV/nucleon | | 400 kW | $10^9$ mrem/hr |

A beam instrumentation is an essential element to successfully operate an accelerator machine in which various diagnostic and beam control systems are integrated. A beam monitor is used to characterize the beam by measuring a spatial distribution of charged particles (e.g. a beam centroid position, angle and their profile), an integrated charge of particles passing through the monitor (e.g. a total beam intensity per spill), and a time structure of charged particles (e.g. a differentiated beam intensity or a bunch structure) along the beam line. If a particle-production target is installed in the beam line, an in-situ target health monitor is a primary sensor to diagnose the target condition and to predict the operational lifetime of the target. When these monitor systems detect any anomaly events, the event signal is analyzed in the system and a proper beam control command is fed into a beam control system (e.g. control a trim magnet) to adjust the beam parameter back to the normal condition. Or, in some cases, the beam is aborted as quick as possible to avoid any catastrophic damages on the accelerator machine. However, the beam instrumentation performance is often constrained by a prompt radiation dose, integrated radiation dose, operation (ambient) temperature and humidity, available space, and strength of embedded electromagnetic fields at the monitor. These constraints will limit the dynamic range of operational beam parameters, like the maximum achievable beam power. Therefore, a seamless R&D effort to develop the radiation hardened beam instrumentations is crucial for making future multi-MW beam facilities.

The most critical element for making the radiation robust monitor is a light guide (e.g. an optical fiber) and an optical element (e.g. a lens and mirror) which are widely used for various optical measurements as beam and target health monitors. These light optics are used to inject a probe light or extract an optical measurement light from a beam line or a target. It is challenging to mitigate the radiation-induced attenuation (RIA) which indicates a lifetime of an optical material as a function of an integrated radiation dose. Because the optical measurement light transmits the beam-related information, a refractive index profile ($\varepsilon$ vs $\lambda$) of the fiber as a function of the integrated radiation dose should be crucial. A high purity silica and Fluoride doped optical material appear higher radiation tolerance than a standard silica optics [2]. Irradiation tests of several radiation resistant fibers were reported in the workshop. Further R&D efforts should be continued. It is also crucial to develop a radiation hardened image sensor. Several groups report that they stop using a CCD camera. J-PARC neutrino group continues to develop a CID camera for the Optical Transition Radiation system. The collider detector group has demonstrated a radiation hardened CMOS which performs in the interaction region. The spatial resolution and data acquisition speed of the CMOS camera are extremely higher than the CCD, while the radiation tolerance on the CCD is an order of magnitude higher than the CMOS. We propose continuous efforts for developing the radiation robust CMOS. Based on the experimental results at the SNS injection area, it seems the major hurdle is the signal processing electronics rather than the sensor itself.

CERN and Fermilab collider groups show the advanced technology for the radiation resistive Application-Specific Integrated Circuit (ASIC) to readout and manipulate a signal from a high-granularity fast detector in an extreme environment for the HL-LHC collider detector and future collider detectors [3]. As an example of the radiation hardened ASIC, a beam halo detector has been demonstrated in LHC. The detector and readout electronics is a unit structure so that it is exposed by an extremely high radiation dose. These technologies will be adopted for making a new type of a fast beam monitor system.

Various types of in-situ target health monitors were reviewed in the workshop. Fiber-optic strain sensors are used to measure mega-hertz dynamic strains on the target vessel induced by individual proton pulses. Acoustic sensor and Laser Doppler Vibrometer (LDV) are used to measure a specific vibration mode in the target system which is generated by the beam impact on the target system. A thermocouple sensor and Infra-Red (IR) detector are used to measure a thermal distribution in the target system. A luminescence light from a Cr doped Alumina is widely used to measure the beam profile on the target. The biggest advantage is that it can cover a large area for a large spot size beam and measure the 2D profile accurately even at the beam tail. Mitigating the aging effect is a key R&D for this. Because the LDV, IR detector, and luminescence monitors measure an optical signal, the radiation hardened light optics is crucial to improve the in-situ target health monitors.

A beam monitoring system to measure primary proton beam parameters were reviewed in the workshop. An Optical Transition Radiation (OTR) monitor and Beam Induced Fluorescence (BIF) monitor have been demonstrated at various beam facilities. Those are a promising technology to apply for a multi-MW beam line because the beam is less distorted by them than other beam profile monitors. Finding a light yield and a stress analysis are a key R&D. Because a small amount of light transmits with the beam related information, maintaining high light transmission efficiency is a key element for the system performance. Secondary Electron emission Monitor (SEM) is other potential candidate for intense beam facilities. It requires an R&D to investigate the aging effect on the Secondary Emission Yield (SEY) and to challenge making a thin wire or film to minimize the interruption of beam. A new SEY material, like Ni, carbon graphite, and Carbon Nano Tube (CNT), are considered for multi-MW accelerators. A new radiation hardened Beam Loss Monitor (BLM) system has been developed at CERN for HL-LHC. There was a comment on a high accuracy beam intensity monitor for neutrino factories. ESS group demonstrates a modulated beam instrumentation which is capable to change out the device quickly and smoothly. This will be a main mechanical structure for the future beam instrumentation.

Secondary and tertiary beam profile monitor were reviewed in the workshop. SEM and gas-RF monitor [4] are considered for a hadron monitor. Electron Multiplier Tube (EMT) [5] and Current Transformer (CT) are a new candidate and are demonstrated at J-PARC for a muon monitor. They appear a good stability with respect to the beam intensity. A continuous R&D is needed for multi-MW beam applications. A pico second muon monitor is considered by the Fermilab group which has a potential to reconstruct the pion phase space in the target system. The proof-of-principle test will be proposed.

Several groups report that the Machine Learning (ML) algorithm is useful for the diagnostic system. The ML is a powerful tool to find a hidden correlation among a vast variance parameter phase and to predict a specific physics parameter. The ML concept is validated with various accelerator applications. The CERN beam monitor group demonstrated reconstructing an optical image emitted from a radiation damaged fiber by using the ML. The Fermilab target group demonstrated the beam parameter prediction by using the muon monitor signal. The result suggests that the ML has a potential to reconstruct the initial pion and kaon phase space by using beam instrumentations. Several groups report that the ML can be used in a beam control feedback loop. The ML effort should be continued for future multi-MW beam facilities.

Table 2: Critical item and their radiation hardness test result reported in the workshop.

| Item | Reported radiation hardness in the workshop | Possible R&D institutions |
|---|---|---|
| Light optics | SNS single-mode fiber sensors: >4×10$^8$ Gy<br>SNS epoxy (Stycast 2850FT with Catalyst 11): > 10$^8$ Gy<br>FRIB viewport: Zinc selenide, no detectable RIA @430 Gy<br>FRIB metal mirror: typically 10$^8$ Gy<br>MLF Ni: 1MWh and found corrosion<br>MLF Au: More accumulate than Ni but found no corrosion<br>CERN rad hard fiber: FIGR-10, tested RIA @694 kGy by using $^{60}$Co [6] (activity is currently stopped) | Various institutions |
| ASIC | FNAL CMS ECON-T: demonstrated up to 500 Mrad<br>CERN luminosity monitor FBCM: design goal 250 Mrad & 3.5e15 1 MeV neq. Fluence; tested 10 Mrad [7] | CERN, Fermilab |
| Camera | CERN camera: SYRA camera tested @10 Mrad (100 kGy)<br>CERN Micro-Cameras & Space Exploration SA [8]: tested @11.7 krad & 36 krad<br>J-PARC neutrino: OTR CID camera operated 1 kGy/year for 750 kW beam power | Various institutions |

| Strain detector [9] | Described in light optics | ESS, J-PARC, SNS |
| --- | --- | --- |
| Acoustic sensor [10] | MLF dynamic mic: tested > 4000 MWh | ESS, J-PARC, SNS |
| LDV [11] | MLF tested total 5000 MWh, total dose 5000 Gy | Fermilab, J-PARC |
| Luminescence beam profile | SNS Cr:Al2O3: 100 MW, ~0.1 DPA | SNS |
| IR monitor | Described in light optics | Various institutions |
| Thermocouple sensor | Extremely radiation robust, no specific dose reported | Various institutions |
| OTR [12] | Described in light optics | Various institutions |
| BIF | Described in light optics | J-PARC, Fermilab |
| SEM [13] | ESS APTM: design goal 4 kGy/h, lifetime > 2 years<br>ESS GRID: design goal 4 kGy/h, 5 DPA/year | Various institutions |
| BLM [14] | CERN BLMASIC: > 100 kGy (10 Mrad) | CERN, Fermilab |
| EMT [15] | J-PARC 100 days at 1.3 MW | J-PARC, Fermilab |
| Gas-filled RF | FNAL 3 yrs test at MI abort line, no signal degradation found | Fermilab |
| CT [15] | No specific report in the workshop | J-PARC, Fermilab |

We propose a global R&D effort to develop the radiation hardened beam instrumentations. As the first step, we will organize a regular workshop to update the progress of each R&D item. If possible, we will form a collaboration group to get more support and to manage the team effort more efficiently.

Reference:
[1] https://indico.fnal.gov/event/52543/
[2] Sylvain Girard et al., "Recent advances in radiation-hardened fiber-based technologies for space applications", J. Opt. 20 (2018) 093001.
[3] White papers
[4] K. Yonehara et al., "Radiation robust rf gas beam detector R&D for intensity frontier experiments", FERMILAB-CONF-19-796-AD.
[5] Y. Ashida et al., "A new electron-multiplier-tube-based beam monitor for muon monitoring at the T2K experiment", Prog. Theor. Exp. Phys. 2018, 103H01.
[6] D. Celeste et al., "Development and Test of a Beam Imaging System Based on Radiation Tolerant Optical Fiber Bundles", Proc. IPAC 2019 Melbourne, Australia, May 2019, WEPGW078, https://accelconf.web.cern.ch/ipac2019/doi/JACoW-IPAC2019-WEPGW078.html.
[7] K.J.R. Cormier et al., "Development of the front end amplifier circuit for the ATLAS ITk silicon strip detector", 2021 JINST 16, P07061.
[8] www.microcameras.ch.
[9] B. Qi et al., "Quadrature phase-shifted optical demodulator for low-coherence fiber-optic Fabry-Perot interferometric sensors", Opt. Express Vol. 27, 7319 (2019); Y. Liu et al., "Upgraded fiber-optic sensors system for dynamic strain measurement in Spallation neutron Source", IEEE Sensors J. Vol 21, 26772 (2021).
[10] M. Teshigawara et al., "Gas yield and mechanical property of super insulator (polyimide and polyester) after gamma ray irradiation" written in Japanese, Teion Kogaku, Vol. 41 (2006) 99.
[11] M. Teshigawara et al., "Development of JSNS target vessel diagnosis system using laser Doppler method", J. Nucl. Mater., 398 (2010), 238-243.
[12] S. Bhadra et al., "Optical transition radiation monitor for the T2K experiment", Nucl. Instr. And Meth. A, vol. 703, 45, 2013.
[13] H. Takada et al., "Recent status of the pulsed spallation neutron source at J-PARC", JPS Conf. Proc. 28, 081003 (2020).